\documentclass{article}[11pt]
\usepackage{a4wide}
\usepackage{bm}
\usepackage{bbm}
\usepackage{epsfig,graphics}
\usepackage{amsmath,amssymb,amsfonts}
\usepackage{color}
\usepackage{epstopdf}
\usepackage{slashed}
\usepackage[
colorlinks=true,
linkcolor=black,
breaklinks=true,
urlcolor=blue,
citecolor=green]{hyperref}

\newcommand{\be}{\begin{equation}}
\newcommand{\ee}{\end{equation}}
\newcommand{\ba}{\begin{eqnarray}}
\newcommand{\ea}{\end{eqnarray}}

\begin{document}

\title{Charmonium, exotic hadrons and hadron structure~\footnote{Talk at the International Symposium of ``50 Years Discovery of the J Particle", Beijing, October 20, 2024.}}
\date{\today}

\author{      Bing-Song Zou \footnote{Email address:
      \texttt{zoubs@mail.tsinghua.edu.cn} }
        \\[2mm]
      {\it\small Department of Physics, Tsinghua University,
 Beijing 100084, China}
}

\maketitle

\begin{abstract}
To celebrate the 50th anniversary of the discovery of the $J/\psi$, the first charmonium state observed,
I start with a brief review of major progresses on the QCD inspired quark potential model originated from charmonium spectrum.Then I show the importance of unquenching dynamics, multiquark components and exotic multiquark states for understanding hadron structure and hadron spectrscopy. The $J/\psi$ and charmonium-like states have played an important role in this aspect.

\end{abstract}

\medskip

\newpage

\section{QCD inspired quark potential model originated from charmonium}

In the early 1960s, a ``particle zoo" of hadrons (mesons and baryons) was discovered, with no clear organizing principle.
In the year of 1964, Murray Gell-Mann~\cite{Gell-Mann:1964ewy} and, independently, George Zweig~\cite{Zweig:1964ruk} proposed that all hadrons were not elementary, but were composed of more fundamental, fractionally charged constituents called quarks (Gell-Mann's term) or ``aces." The original model required three types (``flavors") of quarks: up (u), down (d), strange (s). Mesons were bound states of a quark and an antiquark ($q\bar q$), while baryons were three-quark states ($qqq$). To explain the spin statistics of particles like the $\Delta^{++}$, the quarks were assigned an additional, unobserved quantum number later called color. It was a classification scheme and a static constituent model. It brilliantly organized the hadron spectrum but offered no dynamical explanation for the binding force. 

In the year of 1974, the discovery of the $J/\psi$ particle~\cite{E598:1974sol,SLAC-SP-017:1974ind}, a very narrow resonance around 3.1 GeV, was immediately interpreted as a bound state of a new heavy charm quark and its antiquark ($c\bar c$). Because the charm quark is heavy ($\sim 1.5$ GeV), the $c\bar c$ system is non-relativistic. This allowed people to model it using the highly successful framework of non-relativistic quantum mechanics, analogous to positronium ($e^+e^-$) in QED. Inspired by the emerging theory of Quantum Chromodynamics (QCD)~\cite{Gross:1973id,Politzer:1973fx}, the inter-quark potential was phenomenologically constructed from two key QCD features: 1) Asymptotic Freedom: at short distances, quarks interact weakly via one-gluon exchange, leading to a Coulomb-like potential ($-\alpha/r$); 2) Confinement: at large distances, the force does not diminish but remains constant, leading to a linearly rising potential ($+\kappa r$). The resulting potential the Cornell group: $V(r) = - (4/3)(\alpha_s / r) + \kappa r$ is named as Cornell Potential. This simple ``Coulomb + Linear" form was spectacularly successful in explaining the level splittings, masses, and decays of the charmonium family. It provided the first concrete, QCD-motivated dynamical framework for hadron structure, validating quarks as real dynamical entities.

The basic potential model worked well for heavy quarkonia ($c\bar c$, $b\bar b$) but failed for light quarks (u, d, s), which are highly relativistic~\cite{Godfrey:1985xj,Capstick:1986ter}. Two key theoretical concepts were incorporated to address light hadrons:

{\bf A. Chiral Symmetry and the Goldstone Bosons}

In the limit of massless u, d, s quarks, QCD exhibits chiral symmetry. This symmetry is spontaneously broken in the vacuum, giving rise to pseudoscalar mesons ($\pi$, K, $\eta$) as (pseudo-)Goldstone bosons. For light quarks, the confining potential is not the only dominant force. The exchange of Goldstone bosons (pions, etc.) between light quarks generates a long-range, spin-dependent attractive force. This ``chiral quark model" extension~\cite{Manohar:1983md} was crucial for explaining:
the exceptionally low mass of the pion (Goldstone boson), the hyperfine splittings in the light meson and baryon spectrum (e.g., $N$-$\Delta$ splitting), and the pion-cloud contributions to hadron structure and form factors.

{\bf B. Hidden Local Gauge Symmetry and Vector Meson Exchange}

The interactions of the pseudoscalar meson octet (($\pi$, K, $\eta$) can be described by a nonlinear chiral Lagrangian. Bando, et al.~\cite{Bando:1984ej} showed this Lagrangian possesses a ``hidden" local gauge symmetry when vector mesons ($\rho$, $\omega$, $K^*$, $\phi$) are introduced as the gauge bosons of this symmetry. This formalism provides a rigorous way to include vector meson exchange ($\rho$, $\omega$, etc.) as a fundamental part of the inter-quark or inter-hadron force.
 In the extended chiral quark models~\cite{Glozman:1995fu}, it complements the one-gluon exchange (short-range) and chiral boson exchange (long-range) by providing intermediate-range correlations. The $\omega$ meson provides the attractive force between a quark and an antiquark, while repulsive force is generated between two quarks due to its
G parity. It gives a theoretical basis for the short-to-intermediate range attraction and repulsion in the nucleon-nucleon force ($\omega$ exchange provides repulsion, $\rho$ and $\pi$ provide attraction and spin-orbit forces).

The modern ``chiral quark model" or ``constituent quark model" often synthesizes these ideas:
Constituent quarks (dressed by the QCD vacuum) bound by a QCD-inspired confining potential;
short-range residual one-gluon exchange (or instanton-induced forces); 
medium/long-range forces derived from chiral symmetry (pion and other Goldstone boson exchanges) and hidden gauge symmetry (vector meson exchanges).

This framework provides a remarkably effective phenomenological bridge between fundamental QCD and the observed spectrum and properties of light and heavy hadrons, including exotic states discovered recently. The known ground state mesons and baryons of all kinds of quarks as well as the newly observed tetraquark $T_{cc}$~\cite{LHCb:2021vvq} can be beautifully reproduced with the same set of parameters~\cite{He:2023ucd,He:2023gqh}.

\begin{figure}[htp]
	\centering
	\includegraphics[scale=0.52]{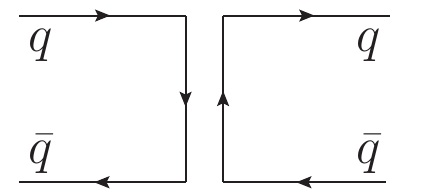}
	\includegraphics[scale=0.52]{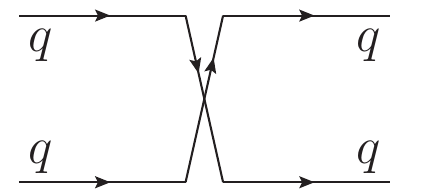}
	\caption{Meson exchange between $q\bar{q}$ (left) and $qq$ (right).
		 }
	\label{fig:feyn_diag}
\end{figure}
While the Cornell potential with only one-gluon exchange potential plus linear confinement potential can reproduce the heavy quarkonium spectrum, it is necessary to introduce effective meson exchange force between light quarks. The fact indicates that the quark-line diagrams as shown in Fig.~\ref{fig:feyn_diag} are important for the light quark interactions.
They cannot be described effectively by one-gluon exchange force.

Although the extended chiral quark model can reproduce the mass spectrum of all known ground state hadrons, it fails for many spatial exited states of various hadrons. This is because such quark model has ignoed unquenching dynamics. 

\section{Penta-quark components in the proton}

The proton is the only stable hadron. It has been discovered over a century. But its internal quark-gluon structure is still not well understood. 
In the classical quark models prior to QCD, each proton was regarded
as composed of three quarks ($uud$). However, evidence has been accumulating for the existence of
significant intrinsic non-perturbative penta-quark components in the
proton. A surprisingly large asymmetry between the $\bar u$ and
$\bar d$ sea quark distributions in the proton with $\bar d-\bar
u\sim 0.12$ was observed from deep inelastic scattering and
Drell-Yan experiments~\cite{Garvey:2001yq}. This demands the $uudd\bar d$ component in the proton to
be more than 12\%. In addition, experiments find non-zero strange quark form factors~\cite{Armstrong:2012bi}, indicating a significant $uuds\bar s$ component. A popular explanation for the
excess of $\bar d$ over $\bar u$ in the proton was given by
meson-cloud model~\cite{Speth:1996pz} by including in the proton a mixture
of $n\pi^+$ with the $\pi^+$ composed of $u\bar d$. An alternative solution is assuming a new configuration of
diquark-diquark-antiquark structure~\cite{Zou:2005xy,An:2005cj}. The diquark cluster
configurations also give a natural explanation for the excess of
$\bar d$ over $\bar u$ in the proton with a mixture of $[ud][ud]\bar
d$ component in the proton. Similarly the asymmetric strangeness distribution in the proton can also be explained by either  $K^+\Lambda$ or $[ud][us]\bar s$ configurations in the proton.

The peta-quark $uudq\bar{q}$ components in the proton with the $\bar q$ of negative intrinsic
parity demand either a quark or anti-quark in the P-wave to make up
the proton of positive intrinsic parity. So it gives naturally the
nonzero quark orbital angular momentum and hence a natural
explanatione~\cite{Zou:2005xy,An:2005cj} to the proton ``spin crisis"~\cite{EuropeanMuon:1987isl} and single spin asymmetry problem~\cite{HERMES:1999ryv}.

To explain the observed $\bar d/bar u$ asymmetry, the strangeness content and spin crisis in the proton consistently, it is  necessary to include about 30\% penta-quark components in the proton. If the proton as the lightest baryon has about 30\% pentaquark components, then one would expect more penta-quark components in its excited states. Therefore it is extremely important to go beyond the conventioanal quenched quark models and to study penta-quark states, to really understand baryon structure.

\section{Exotic hadrons}

The first penta-quark candidate $\Lambda(1405)$ is in fact predicted before the quark model as the $\bar KN$ molecule in 1959 by Daliz and Tuan~\cite{Dalitz:1959dn}, and observed in its decay to $\pi\Sigma$ in 1961~\cite{Alston:1961zzd}.
However after the quark model was proposed in 1964, the $\Lambda(1405)$ resonance was ascribed as an excited
state of a three-quark (uds) system with one quark in an orbital P-wave excitation. Although there were continuously strong arguments on its being hadronic dynamically generated state from chrial dynamics~\cite{Kaiser:1995eg,Oller:2000fj}, 50 years after its prediction and observation as the $\bar KN$ bound state, the PDG in 2010~\cite{ParticleDataGroup:2010dbb} still claimed unambiguously in favor of it as a true three-quark state. Its SU(3) partner, $N^*(1535)$, suffers a similar ambigurity on whether it is a $K\Sigma$-$K\Lambda$-$N\eta$ dynamically generated state~\cite{Kaiser:1995cy} or a three quark state.

Around the end of last century, we proposed to study $N^*$ from charmonium decays~\cite{Zou:2000wg}. For the charmonium decays, each initial state has a single spin and thefixed isospin zero. This simplifies the partial wave analysis and avoids the mixture of $N^*$ and $\Delta^*$ suffered in the $\pi N$ and $\gamma N$ experiments. The study of $N^*$ from $J/\psi$ and $\psi^\prime$ decays observed some new $N^*$ resonances~\cite{BES:2004gwe,BESIII:2012ssm} and shed new light on the properties on the $N^*(1535)$~\cite{BES:2001gvq,Liu:2005pm} and $N^*(1440)$~\cite{Zou:2025nnw}.    

The combined analysis~\cite{Liu:2005pm} of BES data and COSY data supports the $N^*(1535)$ to be a $K\Sigma$-$K\Lambda$-$N\eta$ dynamically generated state with large $uuds\bar s$ component.

Extending from the hidden strangeness to hidden charm naturally led to the expectation for the existence of three narrow  $P_c$ states: one $\bar D\Sigma_c$ bound states and two nearly degenerated $\bar D^*\Sigma_c$ states~\cite{Wu:2010jy, Wu:2010vk, Wang:2011rga, Wu:2012md}, which were suggested to be looked for through their $J/\psi p$  decay mode.  The observation of three narrow $P_c$ states decaying to $J/\psi p$  by LHCb experiment~\cite{LHCb:2015yax,LHCb:2019kea} confirms the pridiction perfectly. 

The isoscalar tetra-quark states with hidden charm of hadronic molecule type were also predicted to exist~\cite{Tornqvist:1993ng,Zhang:2006ix,Gamermann:2006nm}. The narrow charmonium-like state $X(3872)$ decaying to $\pi^+ \pi^- J/\psi$ observed by the Belle Collaboration in 2003~\cite{Belle:2003nnu} seems to fit well the predicted  weakly-bound state of two charm mesons ($D\bar D^*$). But there is no consensus on its fundamental internal structure. Whether it has a large $c\bar c$ component is still under debate~\cite{Chen:2016qju,Guo:2017jvc}.

The isovector tetra-quark states with hidden charm of hadronic molecule type were not expected before the $Z_c(3900)$ was observed through its $\pi J/\psi$ decay mode by BESIII Collaboration in 2013~\cite{BESIII:2013ris}. The attractive force for the isovector $D\bar D^{(*)}$ system is much weaker than for the isoscalar $D\bar D^{(*)}$ system~\cite{Zhang:2006ix}. But it was expected by a $D\bar D^*\to \pi J/\psi$ final state interaction calculation~\cite{Chen:2011xk} which may be enhanced by a triangle singularity~\cite{Wang:2013cya}.

Due to the importance of multi-quark states for understanding the hadron structure and hadron excitation mechanism, the first papers for the dicoveries of $X(3872)$~\cite{Belle:2003nnu}, $Z_c(3900)$~\cite{BESIII:2013ris} and $P_c$ states~\cite{LHCb:2015yax} are top cited ones for the Belle, BESIII and LHCb Collaboration, respectively. Various theoretical explanations for them have been proposed~\cite{Chen:2016qju,Guo:2017jvc}. To distinguish various theoretical pictures for them, the prediction of multi-quark spectra should be provided for each picture to be checked by future experiments. By constructing a nonrelativistic effective field theory with open channels, we discussed the generalities of threshold behavior, and offer an explanation of the abundance of near-threshold peaks in the heavy quarkonium regime~\cite{Dong:2020hxe}. We showed that the threshold cusp can show up as a peak only for channels with attractive interaction, and the width of the cusp is inversely proportional to the reduced mass relevant for the threshold.
With our hadronic molecule framework which predicted successfully the $P_c$ states with vector-meson-exchange dominance~\cite{Wu:2010jy, Wu:2010vk}, we predicted 229 heavy-antiheavy hadronic molecular states~\cite{Dong:2021juy} and about the same amount of heavy-heavy hadronic molecular states~\cite{Dong:2021bvy}.

All the hidden charm hadronic molecules should have their hidden beauty~\cite{Wu:2010rv} and hidden strangeness partners~\cite{Zou:2013af,He:2017aps}. While the two $Z_b$ states observed by Belle Collaboration~\cite{Belle:2011aa} are obviously the beauty partners of the two $Z_c$ states observed by BESIII Collaboration~\cite{BESIII:2013ris,BESIII:2013ouc}, we also found much new evidence for the strange partners of the $P_c$ states by reanalyzing $\gamma p\to\phi p$~\cite{Wu:2023ywu}, $\gamma p\to K^*\Sigma$~\cite{Ben:2023uev}, $\gamma p\to K^{*+}\Lambda$~\cite{Tian:2025bkx} and $\gamma p\to K\Sigma$~\cite{Suo:2025rty}.

While $J/\psi$ played a crucial role for the discoveries of these most famous multiquark states through $X(3872)\to J/\psi\pi\pi$, $Z_c(3900)\to J/\psi\pi$ and $P_c\to J/\psi~p$, it may provide many other interesting results on various possible multiquark states, such as $\eta_1(1855)$ with exotic $J^{PC}=1^{-+}$ quantum numbers observed in $J/\psi\to\gamma\eta\eta'$~\cite{BESIII:2022riz}, which could be explained as a $\bar KK_1(1400)$ molecule~\cite{Dong:2022cuw}. Recently a narrow exotic $0^{--}$ $D^*\bar D_1$ molecule $\psi_0(4360)$ has been predicted and may be searched for through its decay to $J/\psi\eta$~\cite{Ji:2022uie}.

BESIII Collaboration also provided new evidence~\cite{BESIII:2024mbf} in supporting existence of $\Sigma(1380)$~\cite{Wu:2009tu,Wang:2024jyk}, the SU(3) partner of the hadron dynamically generated states $\Lambda(1405$ and $N^*(1535)$.

\section{Unquenched quark model}

With observation of about 30\% penta-quark components in the proton and many multiquark states, it is clear that we must go beyond the coventional quenched quark models by extending it to unquenched quark model to describe the hadron spectrum. The key ingredient for constructing unquenched quark model is the effective quark pair creation mechanism. Among
various quark pair creation mechanisms, the most simple and successful one is the $^3P_0$ model~\cite{LeYaouanc:1972vsx},
where the generated light quark pair share the same quantum numbers as a vacuum.  

\begin{figure}
    \centering
    \includegraphics[width=10cm]{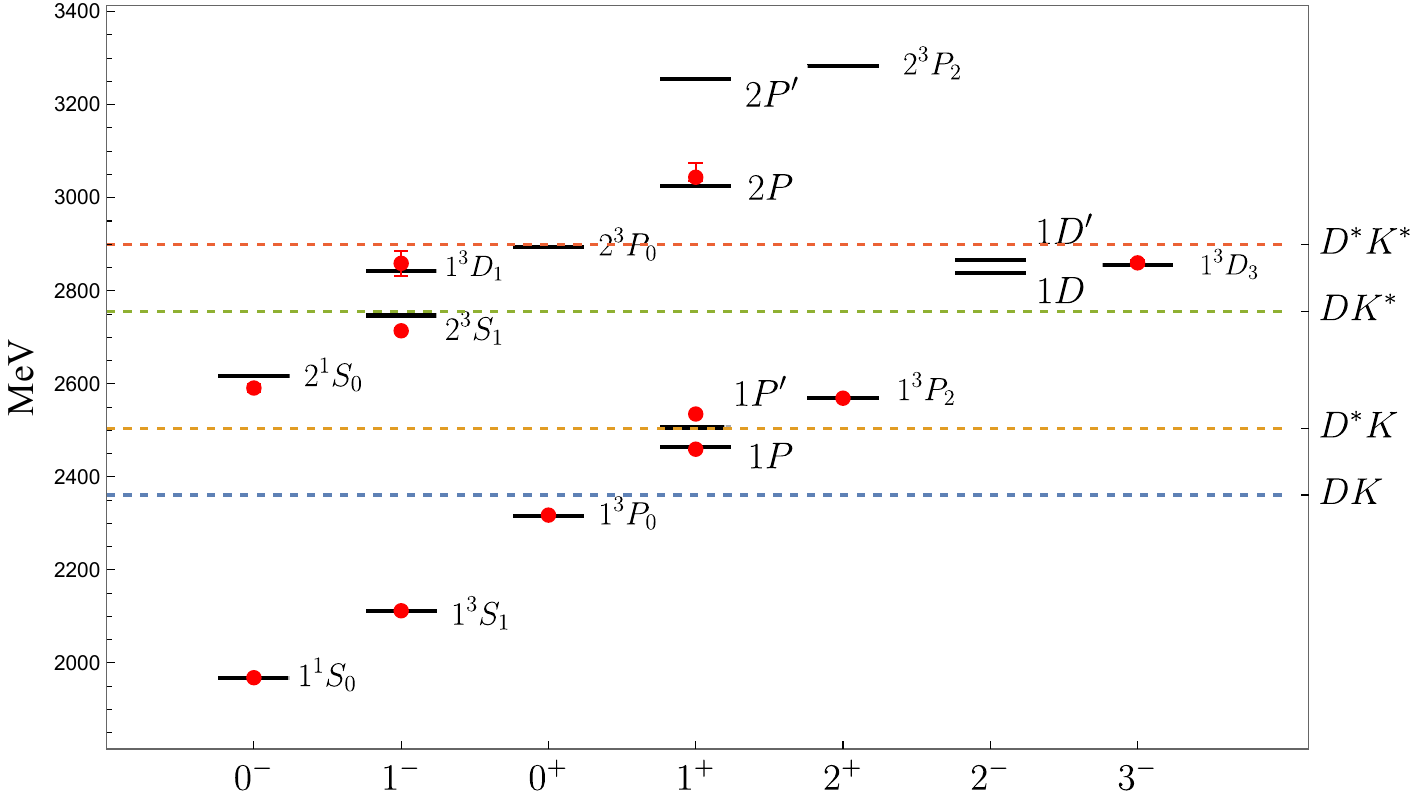}
    \caption{The spectrum of the $D_s$ mesons with the unquenched quark model~\cite{Hao:2022vwt}. Red dots with error bars denote the experimental values from PDG and the model calculations are depicted as black lines.}
    \label{fig:spectrum}
\end{figure}

We have made systematic calculations of the spectra and hadronic decays of the bottomonium~\cite{Lu:2016mbb} and $D_s$ system~\cite{Hao:2022vwt} in a coupled channel framework, where the unquenched effects are induced by the $^3P_0$ model. In the calculation, the
wave functions are obtained by using a nonrelativistic potential model and are handled precisely with
Gaussian expansion method. Even though the fitting mainly focuses on the spectrum, our model agrees
well with the experiments on both the spectra and the hadronic decays, suggesting that the coupled
channel effect could result in a reasonable and coherent description of these meson systems. As an example, the spectrum of the $D_s$ mesons with the unquenched quark model~\cite{Hao:2022vwt} in comparison with data is shown in Fig.\ref{fig:spectrum}.

In the unquenched quark model, it is found that even for the ground state $D_s$ there is 17\% tetra-quark components, a portion camparable with the pentaquark component in the proton. A $D^*K^*$ dominant molecule with mass 2894 MeV is predicted, only 5 MeV below the $D^*K^*$ threshold. 

\section{Summary and prospects}

In summary, the productions and decays of $J/\psi$ have played and will continue to play important roles for hadron spectroscopy. 

A vast array of observed exotic states, including $X(3872)$, charged $Z_c$ and pentaquark $P_c$ states, finds a natural and unified explanation within the hadronic molecule picture. This model, where exotic hadrons are understood as bound composites of standard mesons and baryons, successfully predicts masses near thresholds and specific decay patterns. It provides a powerful organizing principle, suggesting many more such states await discovery at various interaction thresholds.

To fully understand the hadron spectrum, traditional quark models must evolve. The essential step is to "unquench" them by incorporating dynamic mesonic degrees of freedom and coupled-channel effects. This reveals that near-threshold states are not purely compact quark entities but are often dominated by sizable hadronic molecule components. The most accurate descriptions for states like the X(3872) may therefore be mixed state of a compact $c\bar c$  and an extended molecular halo.

Confirming and extending this picture into a complete multiquark spectroscopy requires a multi-pronged experimental effort. Each major facility probes a unique production mechanism to stress-test the nature of these states:
Belle II, BESIII, and STCF act as precision factories in $e^+e^-$ collisions, delivering high-statistics studies of decays and quantum numbers; PANDA at FAIR may utilize $\bar pp$ annihilation to access various quantum numbers;
JLab and EicC may use photoproduction ($\gamma p$) as a critical tool to distinguish compact tetraquarks (larger cross-section) from extended molecules (suppressed cross-section); JPARC may employs $\pi$/K beams to study exotics through formation and their interaction with nuclear matter; LHC Experiments may serve as discovery engines in pp collisions, finding new, heavier states. High energy neutrino beams may act as a new probe to study $uuds\bar s$ component in the proton and look for $uuds\bar c$ penta-quark states~\cite{Qiao:2025bfv}.

The convergence of data from these diverse frontiers will finally allow us to map the landscape of multiquark matter, moving from discovery to definitive structural understanding.


\bibliographystyle{plain}

\begin{thebibliography}{999}

\bibitem{Gell-Mann:1964ewy}
M.~Gell-Mann,
Phys. Lett. \textbf{8} (1964), 214-215
doi:10.1016/S0031-9163(64)92001-3

\bibitem{Zweig:1964ruk}
G.~Zweig,
doi:10.17181/CERN-TH-401

\bibitem{E598:1974sol}
J.~J.~Aubert \textit{et al.} [E598],
Phys. Rev. Lett. \textbf{33} (1974), 1404-1406
doi:10.1103/PhysRevLett.33.1404

\bibitem{SLAC-SP-017:1974ind}
J.~E.~Augustin \textit{et al.} [SLAC-SP-017],
Phys. Rev. Lett. \textbf{33} (1974), 1406-1408
doi:10.1103/PhysRevLett.33.1406

\bibitem{Gross:1973id}
D.~J.~Gross and F.~Wilczek,
Phys. Rev. Lett. \textbf{30} (1973), 1343-1346
doi:10.1103/PhysRevLett.30.1343

\bibitem{Politzer:1973fx}
H.~D.~Politzer,
Phys. Rev. Lett. \textbf{30} (1973), 1346-1349
doi:10.1103/PhysRevLett.30.1346

\bibitem{Godfrey:1985xj}
S.~Godfrey and N.~Isgur,
Phys. Rev. D \textbf{32} (1985), 189-231
doi:10.1103/PhysRevD.32.189

\bibitem{Capstick:1986ter}
S.~Capstick and N.~Isgur,
Phys. Rev. D \textbf{34} (1986) no.9, 2809-2835
doi:10.1103/physrevd.34.2809

\bibitem{Manohar:1983md}
A.~Manohar and H.~Georgi,
Nucl. Phys. B \textbf{234} (1984), 189-212
doi:10.1016/0550-3213(84)90231-1

\bibitem{Bando:1984ej}
M.~Bando, T.~Kugo, S.~Uehara, K.~Yamawaki and T.~Yanagida,
Phys. Rev. Lett. \textbf{54} (1985), 1215
doi:10.1103/PhysRevLett.54.1215

\bibitem{Glozman:1995fu}
L.~Y.~Glozman and D.~O.~Riska,
Phys. Rept. \textbf{268} (1996), 263-303
doi:10.1016/0370-1573(95)00062-3
[arXiv:hep-ph/9505422 [hep-ph]].

\bibitem{LHCb:2021vvq}
R.~Aaij \textit{et al.} [LHCb],
Nature Phys. \textbf{18} (2022) no.7, 751-754
doi:10.1038/s41567-022-01614-y
[arXiv:2109.01038 [hep-ex]].

\bibitem{He:2023ucd}
B.~R.~He, M.~Harada and B.~S.~Zou,
Phys. Rev. D \textbf{108} (2023) no.5, 054025
doi:10.1103/PhysRevD.108.054025
[arXiv:2306.03526 [hep-ph]].

\bibitem{He:2023gqh}
B.~R.~He, M.~Harada and B.~S.~Zou,
Eur. Phys. J. C \textbf{83} (2023) no.12, 1159
doi:10.1140/epjc/s10052-023-12338-5
[arXiv:2307.16280 [hep-ph]].

\bibitem{Garvey:2001yq}
G.~T.~Garvey and J.~C.~Peng,
Prog. Part. Nucl. Phys. \textbf{47} (2001), 203-243
doi:10.1016/S0146-6410(01)00155-7
[arXiv:nucl-ex/0109010 [nucl-ex]].

\bibitem{Armstrong:2012bi}
D.~S.~Armstrong and R.~D.~McKeown,
Ann. Rev. Nucl. Part. Sci. \textbf{62} (2012), 337-359
doi:10.1146/annurev-nucl-102010-130419
[arXiv:1207.5238 [nucl-ex]].

\bibitem{Speth:1996pz}
J.~Speth and A.~W.~Thomas,
Adv. Nucl. Phys. \textbf{24} (1997), 83-149
doi:10.1007/0-306-47073-X{\_}2

\bibitem{Zou:2005xy}
B.~S.~Zou and D.~O.~Riska,
Phys. Rev. Lett. \textbf{95} (2005), 072001
doi:10.1103/PhysRevLett.95.072001
[arXiv:hep-ph/0502225 [hep-ph]].

\bibitem{An:2005cj}
C.~S.~An, D.~O.~Riska and B.~S.~Zou,
Phys. Rev. C \textbf{73} (2006), 035207
doi:10.1103/PhysRevC.73.035207
[arXiv:hep-ph/0511223 [hep-ph]].

\bibitem{EuropeanMuon:1987isl}
J.~Ashman \textit{et al.} [European Muon],
Phys. Lett. B \textbf{206} (1988), 364
doi:10.1016/0370-2693(88)91523-7

\bibitem{HERMES:1999ryv}
A.~Airapetian \textit{et al.} [HERMES],
Phys. Rev. Lett. \textbf{84} (2000), 4047-4051
doi:10.1103/PhysRevLett.84.4047
[arXiv:hep-ex/9910062 [hep-ex]].

\bibitem{Dalitz:1959dn}
R.~H.~Dalitz and S.~F.~Tuan,
Phys. Rev. Lett. \textbf{2} (1959), 425-428
doi:10.1103/PhysRevLett.2.425

\bibitem{Alston:1961zzd}
M.~H.~Alston, L.~W.~Alvarez, P.~Eberhard, M.~L.~Good, W.~Graziano, H.~K.~Ticho and S.~G.~Wojcicki,
Phys. Rev. Lett. \textbf{6} (1961), 698-702
doi:10.1103/PhysRevLett.6.698

\bibitem{Kaiser:1995eg}
N.~Kaiser, P.~B.~Siegel and W.~Weise,
Nucl. Phys. A \textbf{594} (1995), 325-345
doi:10.1016/0375-9474(95)00362-5
[arXiv:nucl-th/9505043 [nucl-th]].

\bibitem{Oller:2000fj}
J.~A.~Oller and U.~G.~Meissner,
Phys. Lett. B \textbf{500} (2001), 263-272
doi:10.1016/S0370-2693(01)00078-8
[arXiv:hep-ph/0011146 [hep-ph]].

\bibitem{ParticleDataGroup:2010dbb}
K.~Nakamura \textit{et al.} [Particle Data Group],
J. Phys. G \textbf{37} (2010), 075021
doi:10.1088/0954-3899/37/7A/075021

\bibitem{Kaiser:1995cy}
N.~Kaiser, P.~B.~Siegel and W.~Weise,
Phys. Lett. B \textbf{362} (1995), 23-28
doi:10.1016/0370-2693(95)01203-3
[arXiv:nucl-th/9507036 [nucl-th]].

\bibitem{Zou:2000wg}
B.~S.~Zou,
Nucl. Phys. A \textbf{684} (2001), 330-332
doi:10.1016/S0375-9474(01)00433-X
[arXiv:hep-ph/0006039 [hep-ph]].

\bibitem{BES:2004gwe}
M.~Ablikim \textit{et al.} [BES],
Phys. Rev. Lett. \textbf{97} (2006), 062001
doi:10.1103/PhysRevLett.97.062001
[arXiv:hep-ex/0405030 [hep-ex]].

\bibitem{BESIII:2012ssm}
M.~Ablikim \textit{et al.} [BESIII],
Phys. Rev. Lett. \textbf{110} (2013) no.2, 022001
doi:10.1103/PhysRevLett.110.022001
[arXiv:1207.0223 [hep-ex]].

\bibitem{BES:2001gvq}
J.~Z.~Bai \textit{et al.} [BES],
Phys. Lett. B \textbf{510} (2001), 75-82
doi:10.1016/S0370-2693(01)00605-0
[arXiv:hep-ex/0105011 [hep-ex]].

\bibitem{Liu:2005pm}
B.~C.~Liu and B.~S.~Zou,
Phys. Rev. Lett. \textbf{96} (2006) no.4, 042002
doi:10.1103/PhysRevLett.96.042002
[arXiv:nucl-th/0503069 [nucl-th]].

\bibitem{Zou:2025nnw}
B.~S.~Zou,
[arXiv:2509.11290 [hep-ph]].

\bibitem{Wu:2010jy}
J.~J.~Wu, R.~Molina, E.~Oset and B.~S.~Zou,
Phys. Rev. Lett. \textbf{105} (2010), 232001
doi:10.1103/PhysRevLett.105.232001
[arXiv:1007.0573 [nucl-th]].

\bibitem{Wu:2010vk}
J.~J.~Wu, R.~Molina, E.~Oset and B.~S.~Zou,
Phys. Rev. C \textbf{84} (2011), 015202
doi:10.1103/PhysRevC.84.015202
[arXiv:1011.2399 [nucl-th]].

\bibitem{Wang:2011rga}
W.~L.~Wang, F.~Huang, Z.~Y.~Zhang and B.~S.~Zou,
Phys. Rev. C \textbf{84} (2011), 015203
doi:10.1103/PhysRevC.84.015203
[arXiv:1101.0453 [nucl-th]].

\bibitem{Wu:2012md}
J.~J.~Wu, T.~S.~H.~Lee and B.~S.~Zou,
Phys. Rev. C \textbf{85} (2012), 044002
doi:10.1103/PhysRevC.85.044002
[arXiv:1202.1036 [nucl-th]].

\bibitem{LHCb:2015yax}
R.~Aaij \textit{et al.} [LHCb],
Phys. Rev. Lett. \textbf{115} (2015), 072001
doi:10.1103/PhysRevLett.115.072001
[arXiv:1507.03414 [hep-ex]].

\bibitem{LHCb:2019kea}
R.~Aaij \textit{et al.} [LHCb],
Phys. Rev. Lett. \textbf{122} (2019) no.22, 222001
doi:10.1103/PhysRevLett.122.222001
[arXiv:1904.03947 [hep-ex]].

\bibitem{Tornqvist:1993ng}
N.~A.~Tornqvist,
Z. Phys. C \textbf{61} (1994), 525-537
doi:10.1007/BF01413192
[arXiv:hep-ph/9310247 [hep-ph]].

\bibitem{Zhang:2006ix}
Y.~J.~Zhang, H.~C.~Chiang, P.~N.~Shen and B.~S.~Zou,
Phys. Rev. D \textbf{74} (2006), 014013
doi:10.1103/PhysRevD.74.014013
[arXiv:hep-ph/0604271 [hep-ph]].

\bibitem{Gamermann:2006nm}
D.~Gamermann, E.~Oset, D.~Strottman and M.~J.~Vicente Vacas,
Phys. Rev. D \textbf{76} (2007), 074016
doi:10.1103/PhysRevD.76.074016
[arXiv:hep-ph/0612179 [hep-ph]].

\bibitem{Belle:2003nnu}
S.~K.~Choi \textit{et al.} [Belle],
Phys. Rev. Lett. \textbf{91} (2003), 262001
doi:10.1103/PhysRevLett.91.262001
[arXiv:hep-ex/0309032 [hep-ex]].

\bibitem{Chen:2016qju}
H.~X.~Chen, W.~Chen, X.~Liu and S.~L.~Zhu,
Phys. Rept. \textbf{639} (2016), 1-121
doi:10.1016/j.physrep.2016.05.004
[arXiv:1601.02092 [hep-ph]].

\bibitem{BESIII:2013ris}
M.~Ablikim \textit{et al.} [BESIII],
Phys. Rev. Lett. \textbf{110} (2013), 252001
doi:10.1103/PhysRevLett.110.252001
[arXiv:1303.5949 [hep-ex]].

\bibitem{Guo:2017jvc}
F.~K.~Guo, C.~Hanhart, U.~G.~Mei{\ss}ner, Q.~Wang, Q.~Zhao and B.~S.~Zou,
Rev. Mod. Phys. \textbf{90} (2018) no.1, 015004
[erratum: Rev. Mod. Phys. \textbf{94} (2022) no.2, 029901]
doi:10.1103/RevModPhys.90.015004
[arXiv:1705.00141 [hep-ph]].

\bibitem{Chen:2011xk}
D.~Y.~Chen and X.~Liu,
Phys. Rev. D \textbf{84} (2011), 034032
doi:10.1103/PhysRevD.84.034032
[arXiv:1106.5290 [hep-ph]].

\bibitem{Wang:2013cya}
Q.~Wang, C.~Hanhart and Q.~Zhao,
Phys. Rev. Lett. \textbf{111} (2013) no.13, 132003
doi:10.1103/PhysRevLett.111.132003
[arXiv:1303.6355 [hep-ph]].

\bibitem{Dong:2020hxe}
X.~K.~Dong, F.~K.~Guo and B.~S.~Zou,
Phys. Rev. Lett. \textbf{126} (2021) no.15, 152001
doi:10.1103/PhysRevLett.126.152001
[arXiv:2011.14517 [hep-ph]].

\bibitem{Dong:2021juy}
X.~K.~Dong, F.~K.~Guo and B.~S.~Zou,
Progr. Phys. \textbf{41} (2021), 65-93
doi:10.13725/j.cnki.pip.2021.02.001
[arXiv:2101.01021 [hep-ph]].

\bibitem{Dong:2021bvy}
X.~K.~Dong, F.~K.~Guo and B.~S.~Zou,
Commun. Theor. Phys. \textbf{73} (2021) no.12, 125201
doi:10.1088/1572-9494/ac27a2
[arXiv:2108.02673 [hep-ph]].

\bibitem{Wu:2010rv}
J.~J.~Wu, L.~Zhao and B.~S.~Zou,
Phys. Lett. B \textbf{709} (2012), 70-76
doi:10.1016/j.physletb.2012.01.068
[arXiv:1011.5743 [hep-ph]].

\bibitem{Zou:2013af}
B.~S.~Zou,
Nucl. Phys. A \textbf{914} (2013), 454-460
doi:10.1016/j.nuclphysa.2013.01.001
[arXiv:1301.1128 [hep-ph]].

\bibitem{He:2017aps}
J.~He,
Phys. Rev. D \textbf{95} (2017) no.7, 074031
doi:10.1103/PhysRevD.95.074031
[arXiv:1701.03738 [hep-ph]].

\bibitem{Belle:2011aa}
A.~Bondar \textit{et al.} [Belle],
Phys. Rev. Lett. \textbf{108} (2012), 122001
doi:10.1103/PhysRevLett.108.122001
[arXiv:1110.2251 [hep-ex]].

\bibitem{BESIII:2013ouc}
M.~Ablikim \textit{et al.} [BESIII],
Phys. Rev. Lett. \textbf{111} (2013) no.24, 242001
doi:10.1103/PhysRevLett.111.242001
[arXiv:1309.1896 [hep-ex]].

\bibitem{Wu:2023ywu}
S.~M.~Wu, F.~Wang and B.~S.~Zou,
Phys. Rev. C \textbf{108} (2023) no.4, 045201
doi:10.1103/PhysRevC.108.045201
[arXiv:2306.15385 [hep-ph]].

\bibitem{Ben:2023uev}
D.~Ben, A.~C.~Wang, F.~Huang and B.~S.~Zou,
Phys. Rev. C \textbf{108} (2023) no.6, 065201
doi:10.1103/PhysRevC.108.065201
[arXiv:2302.14308 [nucl-th]].

\bibitem{Tian:2025bkx}
W.~Y.~Tian, N.~C.~Wei, Y.~F.~Wang, F.~Huang and B.~S.~Zou,
Phys. Rev. C \textbf{112} (2025) no.5, 055203
doi:10.1103/qp7h-tshk
[arXiv:2510.03673 [hep-ph]].

\bibitem{Suo:2025rty}
J.~C.~Suo, D.~Ben and B.~S.~Zou,
Phys. Rev. C \textbf{112} (2025) no.6, 065208
doi:10.1103/ckt7-58sd
[arXiv:2504.05811 [nucl-th]].

\bibitem{BESIII:2022riz}
M.~Ablikim \textit{et al.} [BESIII],
Phys. Rev. Lett. \textbf{129} (2022) no.19, 192002
[erratum: Phys. Rev. Lett. \textbf{130} (2023) no.15, 159901]
doi:10.1103/PhysRevLett.129.192002
[arXiv:2202.00621 [hep-ex]].

\bibitem{Dong:2022cuw}
X.~K.~Dong, Y.~H.~Lin and B.~S.~Zou,
Sci. China Phys. Mech. Astron. \textbf{65} (2022) no.6, 261011
doi:10.1007/s11433-022-1887-5
[arXiv:2202.00863 [hep-ph]].

\bibitem{Ji:2022uie}
T.~Ji, X.~K.~Dong, M.~Albaladejo, M.~L.~Du, F.~K.~Guo and J.~Nieves,
Phys. Rev. D \textbf{106} (2022) no.9, 094002
doi:10.1103/PhysRevD.106.094002
[arXiv:2207.08563 [hep-ph]].

\bibitem{BESIII:2024mbf}
M.~Ablikim \textit{et al.} [BESIII],
Phys. Rev. Lett. \textbf{134} (2025) no.2, 021901
doi:10.1103/PhysRevLett.134.021901
[arXiv:2407.12270 [hep-ex]].


\bibitem{Wu:2009tu}
J.~J.~Wu, S.~Dulat and B.~S.~Zou,
Phys. Rev. D \textbf{80} (2009), 017503
doi:10.1103/PhysRevD.80.017503
[arXiv:0906.3950 [hep-ph]].

\bibitem{Wang:2024jyk}
E.~Wang, L.~S.~Geng, J.~J.~Wu, J.~J.~Xie and B.~S.~Zou,
Chin. Phys. Lett. \textbf{41} (2024) no.10, 101401
doi:10.1088/0256-307X/41/10/101401
[arXiv:2406.07839 [hep-ph]].

\bibitem{LeYaouanc:1972vsx}
A.~Le Yaouanc, L.~Oliver, O.~Pene and J.~C.~Raynal,
Phys. Rev. D \textbf{8} (1973), 2223-2234
doi:10.1103/PhysRevD.8.2223

\bibitem{Lu:2016mbb}
Y.~Lu, M.~N.~Anwar and B.~S.~Zou,
Phys. Rev. D \textbf{94} (2016) no.3, 034021
doi:10.1103/PhysRevD.94.034021
[arXiv:1606.06927 [hep-ph]].

\bibitem{Hao:2022vwt}
W.~Hao, Y.~Lu and B.~S.~Zou,
Phys. Rev. D \textbf{106} (2022) no.7, 074014
doi:10.1103/PhysRevD.106.074014
[arXiv:2208.10915 [hep-ph]].

\bibitem{Qiao:2025bfv}
K.~S.~Qiao and B.~S.~Zou,
[arXiv:2510.23370 [hep-ph]].

\end{thebibliography}

\end{document}